\documentclass[proceedings]{JHEP3} 

\usepackage{epsf,multicol}                      

\newbox\mybox
\newcommand\fverb{\setbox\mybox=\hbox\bgroup\verb}
\newcommand\fverbdo{\egroup\medskip\noindent\fbox{\unhbox\mybox}\ }
\newcommand\fverbit{\egroup\item[\fbox{\unhbox\mybox}]}

\title{B Physics Overview: 
       Rare Hadronic and Radiative Decays}

\author{M. Beneke\thanks{Talk presented at the International 
Conference on CP Violation (KAON 2001), Pisa, Italy, 12 June - 17 June, 2001}\\
        Institut f\"ur Theoretische Physik E, RWTH Aachen,\\ 
        Sommerfeldstr. 28, D - 52074 Aachen\\
        E-mail: \email{mbeneke@physik.rwth-aachen.de}}

\conference{PITHA 01/08, hep-ph/0109243, 25 September 2001}

\abstract{I discuss three $B$ physics results with potential 
for exhibiting new flavour-changing interactions: (1) the time-dependent 
CP asymmetry in $B\to J/\psi K_S$ decay; (2) branching fractions and 
CP asymmetries in $B\to \pi\pi,\,\pi K$ decays; (3) inclusive and 
exclusive semi-hadronic $b\to s$ transitions.}

\begin{document} 

\section{Introduction}

The primary motivation for $B$ physics is to complete the determination 
of the CKM matrix including its CP violating phase and to perhaps 
discover flavour-changing interactions beyond the standard weak 
interaction. With the large number of $B$ mesons now being produced 
at SLAC and KEK, these efforts have entered a new phase in which the 
standard theory will finally be seriously challenged. For this talk 
I have chosen to discuss the theory and interpretation of three sets 
of key observables, for which experimental results are already
available, and are likely to be much improved 
in the near future. 

\section{Time-dependent CP asymmetry in $B\to J/\psi K_S$ decay}

The time-dependent asymmetry in $B\to J/\psi K_S$ decay is predicted 
to be \cite{Bigi:1981qs,Dunietz:1986vi} 
\begin{eqnarray}
\label{mixphase}
A_{\rm mix}(t) &\equiv& 
\frac{\Gamma(\bar B^0(t)\to J/\psi K_S)-\Gamma(B^0(t)\to J/\psi K_S)}
{\Gamma(\bar B^0(t)\to J/\psi K_S)+\Gamma(B^0(t)\to J/\psi K_S)} 
\nonumber\\[0.2cm]
&=& \sin \Phi_d \sin(\Delta M_{B_d} t),
\end{eqnarray}
i.e. it measures the phase of the $B\bar B$ mixing amplitude,
$\Phi_d$, assuming the standard parameterization of the CKM matrix 
in which $V_{cb} V_{cs}^*$ has (almost) no phase. In the Standard 
Model $\Phi_d=2\beta$ (with $\beta$ one of the angles of the 
unitarity triangle) is related to the phase of $V_{td}$.  

Eq.~(\ref{mixphase}) is based on several assumptions, all of which are
verified experimentally or expected to be valid 
to the percent level or better: 
\begin{itemize}
\item[-] The lifetime difference of the $B_d$ mass eigenstates and 
CP violation in the $B\bar B$ mixing amplitude is negligible.
\item[-] CP violation in $K \bar K$ mixing is negligible.
\item[-] The decay proceeds only through amplitudes with a common 
weak phase, the phase of $V_{cb} V_{cs}^*$. 
\end{itemize}
The last assumption can be checked by fitting a 
$\cos (\Delta M_{B_d} t)$ term to the time-dependence of $A_{\rm mix}(t)$, 
although the absence of such a term 
does not rigorously exclude a decay amplitude with a different weak 
phase. The above assumptions continue to hold in reasonable extensions
of the standard model, so that the CP asymmetry $A_{\rm mix}(t)$ 
always measures the 
$B\bar B$ mixing phase, but it may or may not equal $2\beta$. 

The current results reported by BaBar ($0.59\pm
0.15$ \cite{Aubert:2001nu}) 
and Belle ($0.99\pm 0.15$ \cite{Abe:2001xe}) 
together with an earlier CDF measurement \cite{Affolder:2000gg}
average to\footnote{This has been updated to account for new data 
published after the conference. At the time of the conference the 
result was $0.47\pm 0.16$.} 
\begin{equation}
\label{s2b}
\sin(2\beta) = 0.79\pm 0.10.
\end{equation}
This is in good agreement with an indirect determination of the 
$B\bar B$ mixing phase that follows from the mass difference of 
the $B_d$ and $B_s$ eigenstates, CP violation in $K\bar K$ mixing and 
$|V_{ub}/V_{cb}|$: $\sin(2\beta) = 0.68\pm 0.21$ \cite{Hocker:2001xe}, 
$0.68\pm 0.15$ \cite{Ciuchini:2001de} at $95\%$CL. 

A large difference between the direct and indirect determination 
of the $B\bar B$ mixing phase would probably have implied a new source of 
flavour violation beyond the CKM matrix. Because of this the good
agreement is perhaps not very surprising, since models with non-CKM 
flavour violation are generally disfavoured by the smallness of 
CP violation in $K\bar K$ mixing and the suppression of
flavour-changing neutral currents. Eq.~(\ref{s2b}) therefore 
implies two important facts about Nature:
\begin{itemize}
\item[1)] The Kobayashi-Maskawa mechanism \cite{Kobayashi:1973fv} 
is most likely the dominant source of CP violation at the electroweak 
scale.
\item[2)] CP violating phases are large, i.e. 
CP is not an approximate symmetry.\footnote{In extensions 
of the Standard Model the observed amount of CP violation in the 
kaon system can be explained with small CP violating phases avoiding 
the requirement of a large CKM phase. This possibility might have been
attractive from a conceptual point of view, since most extensions 
of the SM cause too much CP violation with generic CP violating
phases. (However, the problem of large flavour-changing neutral 
currents would then have had to be solved by other means.)}
\end{itemize}

\section{Branching fractions and 
CP asymmetries in $B\to \pi\pi,\,\pi K$ decays}

The decay $B\to J/\psi K$ with its unambiguous theoretical
interpretation is unique. The generic situation is that $B$ decays 
which are interesting for constraining the unitarity triangle 
have significant decay amplitudes with different CP violating 
phases. This leads to new CP violating observables (``direct 
CP violation'') but to determine the CP phases one needs knowledge 
of the relative magnitudes of the different amplitudes -- a
strong interaction problem. There exists now a continuously increasing
number of measured charmless $B$ decay modes, which are presumed 
to have two (nearly) comparable amplitudes, and which are sensitive 
to relative phases due to interference of the amplitudes. 
In the following I discuss
methods to extract information on the phase of $V_{ub}$ from 
decays to pions and kaons. These final states have attracted 
interest recently, since the branching fractions for some 
decays into pions are observed to be 
somewhat lower than expected, while those 
into a pion and a kaon are larger than expected.

Two complementary strategies have been adopted to approach the 
strong interaction problem. The first approach begins with a general 
parameterisation of the strong interaction amplitudes based on 
SU(2) isospin symmetry. One then eliminates as many as possible 
strong interaction parameters through measurements. Additional, 
but less accurate constraints may follow from assuming SU(3) 
flavour symmetry. This approach is limited by the need of many 
accurate measurements and by SU(3) breaking effects. 
The second approach is based on a calculation 
of the strong interaction amplitudes. The feasibility of 
such calculations has only recently been understood and relies 
on the heavy quark limit. The limitation of this approach is that 
the $b$ quark mass is not that large.

\subsection{Flavour symmetry}

As is well-known the time-dependent CP asymmetry in $B\to \pi^+\pi^-$
decay determines $\gamma$, if the penguin amplitude is
neglected.\footnote{I assume here that the $B\bar{B}$ mixing phase is
  taken from experiment. Then $B\to \pi^+\pi^-$ determines the phase 
of $V_{ub}^*$, i.e. $\gamma$.} However, this is not a good 
approximation. 

Neglecting only electroweak penguin amplitudes the isospin amplitude system 
for the three $\pi\pi$ final states contains five real 
strong interaction parameters, just as many as there are independent 
branching fractions under the same assumption.\footnote{The
  electroweak penguin amplitude is of order $1\%$ and can be corrected
  for analytically \cite{Gronau:1999fn,Buras:1999rb}.} Solving this system 
allows one to determine $\gamma$ up to discrete 
ambiguities \cite{Gronau:1990ka}. Since this method requires 
a measurement of the small $B\to\pi^0\pi^0$ branching fractions, it has 
practical difficulties. Already bounds on the CP averaged $\pi^0\pi^0$
branching fraction can be useful to constrain the amplitude 
system \cite{Grossman:1998jr,Charles:1999qx,Gronau:2001ff}. 
If $\mbox{Br}(\pi^0\pi^0)$ is small, the strong phase of the 
penguin-to-tree ratio cannot be large. In fact
$\mbox{Br}(\pi^0\pi^0)= 0$ implies $\mbox{Br}(\pi^+\pi^-)= 2\,
\mbox{Br}(\pi^\pm\pi^0)$. Conversely, a deviation from the last 
relation implies that $\mbox{Br}(\pi^0\pi^0)$ cannot be too small. At
this moment the $\pi^\pm\pi^0$ measurement is still too uncertain to
draw meaningful conclusions. Further constraints on the $\pi\pi$ 
modes can be obtained only by assuming also SU(3) or U-spin
symmetry. This relates, for example, 
 $B_d\to\pi^+\pi^-$ to $B_s\to K^+ K^-$. The 
inverted CKM hierarchy of penguin and tree amplitude in 
the second decay can in principle 
be used to determine $\gamma$ from a combined
measurement of the time-dependent and direct CP 
asymmetries in both decays \cite{Fleischer:1999pa}.

The $B\to \pi K$ decays are penguin-dominated, because the tree 
amplitudes are CKM suppressed. The final states $\pi^0 K^\pm$ and 
$\pi^0 K^0$ have significant electroweak penguin contributions. The
amplitude system contains 11 real strong interaction parameters, too
many to determine them all by measurements. Flavour symmetry is useful
to constrain some of the amplitude parameters:
\begin{itemize}
\item[-] Isospin symmetry implies \cite{Neubert:1999re}
\begin{equation}
{\displaystyle \mbox{Br}(\pi^0 \bar K^0) = 
\frac{\mbox{Br}(\pi^+ K^-)\mbox{Br}(\pi^- \bar K^0)}
{4 \mbox{Br}(\pi^0 K^-)}}\times\left\{1+O(\epsilon^2)\right\},
\end{equation}
where $\epsilon\sim 0.3$ is related to the tree-to-penguin 
ratio.\footnote{Here and in the remainder of the text ``Br'' 
always refers to CP averaged branching fractions.} 
Unless the correction term is unexpectedly large this relation 
suggests a $\pi^0 K^0$ branching fraction of order $6\times 10^{-6}$, 
about a factor $1.5-2$ smaller than the current measurements. 
Further isospin relations of this type have been 
derived \cite{Matias:2001ch}.
\item[-] SU(3) or U-spin symmetry imply:
\begin{itemize}
\item[a.] The dominant electroweak penguin amplitude is 
determined \cite{Neubert:1998jq}.
\item[b.] The magnitude of the tree amplitude for $I=3/2$ final states
  is related to $\mbox{Br}(\pi^\pm\pi^0)$.
\item[c.] Rescattering and annihilation contributions to 
the (otherwise) pure penguin decay $B^+\to\pi^+K^0$ are 
constrained by  $\mbox{Br}(K^+ K^0)$, where they are CKM enhanced
relative to the penguin amplitude \cite{Falk:1998wc}.
\end{itemize}
\end{itemize}
SU(3) flavour symmetry together with a few further dynamical
assumptions (detailed below) suffice to derive bounds on $\gamma$ 
from CP averaged branching fractions alone. The 
inequality \cite{Fleischer:1998um}
\begin{equation}
\sin^2\gamma \leq {\displaystyle \frac{\tau(B^+)}{\tau(B_d)}\,
\frac{\mbox{Br}(\pi^+ K^-)}{\mbox{Br}(\pi^-\bar{K}^0)}} \equiv R
\end{equation}
excludes $\gamma$ near $90^\circ$ if $R<1$ and is derived upon
assuming that the rescattering contribution mentioned above and 
a colour-suppressed electroweak penguin amplitude are negligible. 
Current data give $R=1.06\pm 0.18$. The ratio of charged decay 
modes satisfies \cite{Neubert:1998jq} 
(neglecting again the rescattering contribution to 
$B^+\to\pi^+K^0$) 
\begin{equation} 
\label{nrbound}
2\cdot {\displaystyle
\frac{\mbox{Br}(\pi^0 K^-)}{\mbox{Br}(\pi^-\bar{K}^0)}} \equiv
R_\ast^{-1} \leq \left(1+\bar{\epsilon}_{3/2} 
\left|q-\cos\gamma\right|\right)^2 
+ \,\bar{\epsilon}_{3/2}^2\sin^2\gamma,
\end{equation}
where $q$ and $\bar \epsilon_{3/2}$ are determined according to a. and b. 
above, respectively. This bound is particularly interesting, since, 
if $R_\ast^{-1}>1$,  
it excludes a region in $\gamma$ around $55^\circ$, which is favoured 
by the indirect unitarity triangle constraints. Current data give 
$R_\ast^{-1}=1.40\pm 0.23$. This prefers $\gamma>90^\circ$, but the 
error is still too large to speculate about the implications of this 
statement. Eq.~(\ref{nrbound}) can be turned into a determination 
of $\gamma$ if one assumes that the strong phase of the tree amplitude
relative to the penguin amplitude is not too
large \cite{Beneke:2001ev}. This assumption is justified by theoretical 
calculations as discussed next, 
but will eventually be verified experimentally by the 
observation of small direct CP asymmetries.

\subsection{QCD factorisation}

QCD calculations of hadronic two-body decays deal with matrix elements 
\begin{equation}
\label{mes}
\langle M_1 M_2|O_i|\bar B\rangle
\end{equation}
of operators $O_i$, which appear in the weak effective Hamiltonian. A
widely used approach has been to approximate the matrix element by the
factorisation ansatz \cite{Bauer:1987bm}, 
but the validity of the approximation
has never been clear, apart from its inadequacy to describe strong 
phases. We have shown recently \cite{Beneke:1999br,Beneke:2000ry} that 
the arguments that lead to perturbative factorisation theorems for
many strong interaction processes at large momentum transfer also 
imply that the matrix elements (\ref{mes}) factor into short-distance
and long-distance parts. The short-distance scale is provided by the 
large mass of the decaying meson and the correspondingly large energy of
the final state mesons. The long-distance parts are sufficiently
simple to make this approach predictive. Schematically, the 
matrix element is computed as 
\begin{equation}
 F_{B\to M_1}\cdot T^I_i\ast 
\Phi_{M_2} + \,\Phi_B\ast T^{II}_i \ast \Phi_{M1}\ast\Phi_{M2}
\end{equation}
in the heavy quark limit, i.e. neglecting corrections that scale 
as $1/m_b$. In this equation $F_{B\to M_1}$ denotes a form factor and 
$\Phi_X$ denote the meson's leading-twist light cone distribution 
amplitudes, which are taken as non-perturbative inputs. The
$T^{I,II}_i$ represent perturbative functions, which can be computed 
as series in $\alpha_s$, the strong coupling.
A dependence on the distribution amplitudes arises only at order 
$\alpha_s$; at leading order the factorisation formula reproduces 
the earlier factorisation ansatz, but without phenomenological 
parameters. Strong interaction phases are generated by perturbative
rescattering in the heavy quark limit and appear through imaginary
parts of the hard-scattering functions $T^{I,II}_i$. For decays to 
pions and kaons these functions have been computed to next-to-leading 
order \cite{Beneke:2001ev,Beneke:1999br,Du:2001ns,Muta:2000ti}. 
Since the factorisation formula holds only in the heavy quark limit
one may expect non-negligible $1/m_b$ corrections. This is true in
particular for strong interaction phases, which are either of order 
$\alpha_s$ or $\Lambda_{\rm QCD}/m_b$. Some potentially large 
corrections have been estimated and have been included in a
theoretical error estimate, but a general parametrisation of $1/m_b$
effects has neither been given nor is it likely to be useful in 
practice. It is
therefore important to devise tests of the theory while simultaneously
extracting information on CP violation. 

The main results of our analysis \cite{Beneke:2001ev} of 
$\pi\pi$ and $\pi K$ final states are briefly summarised
as follows:
\begin{itemize}
\item[-] The branching fractions for the modes $B^\mp\to\pi^\mp\pi^0$
  and $B^\mp\to\pi^\mp\bar K^0$, which depend only on a single weak
  phase to very good approximation, 
  are well described by the theory. This demonstrates that the
  magnitude of the tree and penguin amplitude is obtained
  correctly. There is, however, a relatively large normalisation 
  uncertainty for the $\pi K$ final states, which are sensitive to 
  weak annihilation and the strange quark mass through the scalar
  penguin amplitude. This uncertainty can be partially eliminated by
  taking ratios of branching fractions.
\item[-] Branching fractions with a significant interference of tree
  and penguin amplitudes deviate slightly from their 
  measured values \cite{Cronin-Hennessy:2000kg,Abe:2001nq,Aubert:2001hs}, 
  if $\gamma<90^\circ$. As a consequence charmless $B$ decays appear to
  favour larger values of $\gamma$ than the standard unitarity
  triangle fit, although again the errors are too large to reach a
  definite conclusion.
\item[-] With some exceptions (such as the final state $\pi^0\pi^0$)
  strong phase differences are not large, so that QCD factorisation
  predicts small direct CP asymmetries, up to $(10-15)\%$ (with a 
  preference around $5\%$) for the $\pi^0K^\pm$, $\pi^\pm K^0$ final
  states. The current data seem to favour small 
  CP asymmetries \cite{Aubert:2001hs,Chen:2000hv,Abe:2001hs,Aubert:2001qj}, 
  but they are not yet
  accurate enough to decide upon whether the QCD factorisation approach
  allows one to predict strong phases quantitatively, or whether it 
  is limited to the qualitative statement that strong phases are 
  small. Establishing  
  direct CP violation in $B$ decays will not only constitute an
  important physics result but also provide a crucial test of our
  theoretical framework.
\end{itemize}
\begin{figure}[t]
\epsfxsize=10cm
\centerline{\epsffile{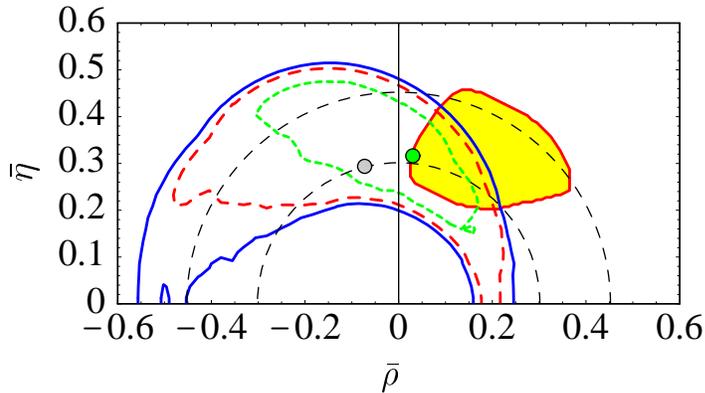}}
\caption{\label{fig:chi2BaBar}\it 
95\% (solid), 90\% (dashed) and 68\% (short-dashed) confidence level 
contours in the $(\bar\rho,\bar\eta)$ plane obtained from a global 
fit to the CP averaged $B\to\pi K,\pi\pi$ branching fractions, using 
the scanning method \cite{Hocker:2001xe}. The darker dot shows the
overall best fit, whereas the light dot indicates the best fit for the
default parameter set.}
\end{figure}
The current interpretation of charmless non-leptonic 
$B$ decays can be summarised
by a global fit of the Wolfenstein parameters $\bar \rho$, $\bar \eta$
to the six CP averaged $\pi\pi$, $\pi K$ branching fractions shown in 
Figure~\ref{fig:chi2BaBar}. The constraint on $V_{ub}$ from
semi-leptonic charmless $B$ decays is not used in this fit. The result
is consistent with the conventional unitarity triangle 
fit \cite{Hocker:2001xe} (shown as the shaded (yellow) region in 
Figure~\ref{fig:chi2BaBar}), but prefers somewhat larger $\gamma$ or
smaller $|V_{ub}/V_{cb}|$. The quality of the fit is very good,
independent of the phenomenological parameter $\rho_A$ which
quantifies the size of weak annihilation. 
The darker (green) dot shows the value of $(\bar\rho,\bar\eta)$ at which
the $\chi^2$ function is minimized within the allowed ranges of
theoretical input parameters. At this minimum
$\chi^2/\mbox{d.o.f.}\approx 0.5$, $\gamma=85^\circ$, 
$|V_{ub}/V_{cb}|=0.071$. The corresponding theory input parameters and
output branching fractions are given in Table~\ref{tab1}.\footnote{
The results discussed here disagree with another 
analysis \cite{Ciuchini:2001gv}
 that has
prematurely declared a failure of QCD factorisation. This 
overly pessimistic  assessment of the current situation is based on an
incomplete use of the QCD factorisation input (neglecting in
particular radiative corrections to the scalar penguin contribution)
together with neglecting errors in some of the important theory input 
parameters. A phenomenological parameter (similar to $\rho_A$ above)
for the QCD penguin amplitude is then 
introduced as a further fit parameter. Despite this, due to the rigid 
theory input parameters, the quality of the fit is in fact worse than
the fit based on QCD factorisation.}

\tabcolsep=0.2cm
\begin{table}[t]
\begin{center}  
\caption{\it \label{tab1}Input and output of the global $(\bar\rho,\bar\eta)$
  fit. CP averaged branching fractions in units of $10^{-6}$.}
\vskip2cm
\hspace*{5cm}
\begin{tabular}{|l|c|c|}
\hline\hline
 Decay Mode & \hspace*{0.1cm}  Fit \hspace*{0.1cm} & 
 Exp. Average\hspace{0.15cm} \\
\hline
$B^0\to\pi^+\pi^-$ & 4.6 & $4.4\pm 0.9$ \\
$B^\pm\to\pi^\pm\pi^0$ & 5.3 & $5.6\pm 1.5$ \\
\hline
$B^0\to\pi^\mp K^\pm$ & 17.9 & $17.2\pm 1.5$ \\
$B^\pm\to\pi^0 K^\pm$ & 11.3 & $12.1\pm 1.7$ \\
$B^\pm\to\pi^\pm K^0$ & 17.7 & $17.2\pm 2.5$ \\
$B^0\to\pi^0 K^0$ & 7.1 & $10.3\pm 2.5$ \\
\hline\hline
\end{tabular}
\end{center}

\vspace*{-5.5cm}

\vspace{0.1cm}
   \begin{center} 
\hspace*{-8cm}
     \begin{tabular}{| r l|} 
\hline  
\hline 
     $F_0^{B\to \pi}(0)$    & 0.312 \\ 
        $f_B$                           & $210$~MeV\\ 
     $|V_{cb}| $            & 0.039 \\
        $\mu$                           & $2.5$~GeV$$ \\
     $m_s(2\,\mbox{GeV})$   & $95$~MeV \\
        $\bar m_c$                      & $1.25$~GeV \\
     $\lambda_B$               & $200$~MeV \\
        $R_{\pi K}$                     & 0.8  \\ 
     $\Lambda_{\rm QCD}$   & $225$~MeV \\
       $a_1(\bar K)$                    & $0.15$ \\ 
     $a_2(\bar K)$   & $0.24$ \\
       $a_2(\pi)$    & $0.32$  \\
     $\varrho_H$                           & 0.4 \\
     $\varrho_A$   & 0 \\[0.1cm]\hline\hline
       \end{tabular} 
\end{center} 
\end{table} 

\section{Inclusive and 
exclusive semi-hadronic $b\to s$ transitions}

The flavour-changing neutral current transition $b\to s\gamma$ has been 
constraining extensions of the standard model severely, since it was 
first observed by CLEO \cite{Ammar:1993sh}. The decay amplitude is not 
only loop-suppressed (a property in common with $B^+\to\pi^+ K^0$ discussed 
above), but also suppressed due to the chiral weak interactions. In the 
effective low-energy theory the transition occurs through the operator 
$g\bar s\sigma^{\mu\nu} (1+\gamma_5) b F_{\mu\nu}$. The 
left-handedness of weak 
interactions implies that the coefficient scales with $m_b/M_W^2$, so 
that the operator is effectively dimension six. New, non-chiral, flavour 
interactions can replace $m_b$ by a weak scale quantity, elevating the 
operator to dimension five.

Because the leading contribution is loop-induced, the computation of the 
first correction in renormalisation group improved perturbation theory 
is a difficult task, which however has been 
completed \cite{Adel:1994ah,Greub:1996jd,Chetyrkin:1997vx}. 
Recently, it has been 
suggested to parametrise the dominant correction in terms of the 
charm quark $\overline{\rm MS}$ mass rather than the pole 
mass \cite{Gambino:2001ew}. This 
increases the theoretical prediction by about $10\%$ to
\begin{equation}
\mbox{Br}(B\to X_s\gamma)_{|E_\gamma>m_b/20} = 
(3.73\pm 0.30)\cdot 10^{-4}
\end{equation}
and suggests that the theory is perhaps not yet as precise as commonly 
assumed. In view of this the prediction seems to be consistent with 
the experimental average \cite{cleobs,bellebs,Barate:1998vz} 
$(2.96\pm 0.35)\cdot 
10^{-4}$. The experimental error includes a theoretical error from 
extrapolating the photon energy spectrum to small photon energies, which 
are not measured.

The exclusive transitions $B\to K^*\gamma$ are easier to measure, but 
more dependent on hadronic parameters. One obstacle to a precise prediction 
of the exclusive decay has recently been 
removed \cite{Beneke:2001at,Bosch:2001gv} by demonstrating that 
non-factorisable strong interaction effects can be computed in 
perturbation theory in the heavy quark limit. The exclusive mode can 
then be treated by renormalisation group methods in a manner analogous 
to the inclusive mode and is now also known to next-to-leading-logarithmic 
order. The theoretical prediction is proportional to the $B\to K^*$ 
tensor form factor $T_1(0)$, and is given by 
\begin{equation}
\mbox{Br}(\bar{B}\to \bar K^*\gamma) =
(7.3\pm 1.4)\cdot 10^{-5}\times\,\left(\frac{\tau_B}{1.6\mbox{ps}}\right) 
\left(\frac{\bar m_b(\bar m_b)}{4.2\,\mbox{GeV}}\right)^{\!2}
\left(\frac{T_1(0)}{0.38}\right)^{\!2}. 
\end{equation}
The next-to-leading order correction is about $30\%$ on the amplitude 
level, enhancing the decay rate significantly. This enhancement is 
not unexpected since a large part of the correction is identical to the 
equally large NLO correction to the inclusive rate. The theoretical 
prediction is now significantly higher than the experimental results 
$\mbox{Br}(\bar{B}^0\to \bar K^{*0}\gamma)_{\rm exp} =
(4.54\pm 0.37)\cdot 10^{-5}$ and 
$\mbox{Br}(B^-\to \bar K^{*-}\gamma)_{\rm exp} = 
(3.81\pm 0.68)\cdot 10^{-5}$ \cite{Coan:2000kh,babarmoriond,bellebs}. 
Given the agreement in the inclusive 
sector a non-Standard Model explanation of this difference appears 
unlikely since it would have to be connected with spectator specific 
interactions. The difference must then be blamed on 
a sizeable $1/m_b$ correction to the amplitude or a smaller 
tensor form factor. Isospin breaking effects are absent at leading order 
in $1/m_b$ in the decay amplitude, so that the branching fractions 
for charged and neutral $B$ mesons may differ only through the different 
$B$ meson lifetimes in this approximation. If the trend indicated by the 
experimental result is real, it may signal a gross failure of the 
theory or a new isospin-violating interaction. In both cases this 
effect should be seen more clearly in $B\to \rho\gamma$ decay. 

Requiring the inclusive branching fraction to be consistent with data 
in the Standard Model extended by a second Higgs doublet implies that 
the charged Higgs boson cannot be 
light \cite{Ciuchini:1998xe,Borzumati:1998tg}. The same is 
true in the supersymmetric standard model unless chargino exchange 
interferes destructively with the charged Higgs contribution. This constrains 
the model parameter space, requiring in particular 
that the product $A_t\mu$ of the soft-supersymmetry breaking parameter 
$A_t$ and the $\mu$-parameter be negative. For models with 
large $\tan\beta$ the theoretical prediction becomes unreliable, unless 
corrections enhanced by powers of $\tan\beta$ are resummed. This 
resummation strengthens the constraints on the parameter 
space \cite{Carena:2001uj,Degrassi:2000qf}. 
The discussion of charged Higgs boson and 
chargino effects in supersymmetry usually proceeds under the assumption 
that flavour-changing gluino vertices are negligible. Once such 
couplings are allowed, implying non-diagonal squark mass matrices in a 
standard quark mass eigenstate basis, they constitute generically 
the dominant supersymmetric effect, so that the data severely limits 
the magnitude of these additional flavour-violating 
interactions \cite{Gabbiani:1996hi}. Including all supersymmetric effects 
simultaneously opens the possibility for cancellations \cite{Besmer:2001tv}, 
but the general conclusion remains true that the allowed squark mass matrix 
pattern is rather restricted.

\begin{figure}[t]
\vspace*{3.4cm}
\hspace*{-1.7cm}
\epsffile{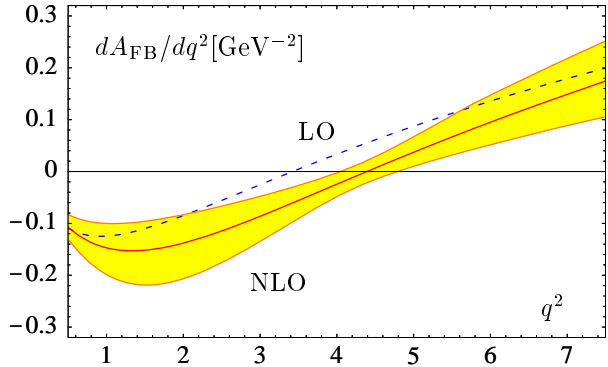}
\vskip0.1cm
\caption{\label{fig:dafbnorm}\it 
Forward-backward asymmetry 
$d A_{\rm FB}(B^-\to K^{*-}\ell^+\ell^-)/dq^2$ 
at next-to-leading order (solid center line) and leading order 
(dashed). The band reflects all theoretical uncertainties from 
parameters and scale dependence combined.}
\end{figure}

The transition $b\to s\ell^+\ell^-$ opens new possibilities to probe 
new flavour interactions, since it is sensitive to flavour-changing 
$Z$ boson couplings $\bar s \Gamma b Z$ \cite{Buchalla:2001sk}. At low energies 
the corresponding dimension-six operators involve two Wilson coefficients, 
$C_9$ and $C_{10}$. Both can be determined 
from the lepton invariant mass spectrum and the forward-backward 
asymmetry. The forward-backward asymmetry is a particularly interesting 
quantity \cite{Ali:1991is}, since the value $q_0$ at which the 
asymmetry vanishes for the exclusive decay $\bar B\to \bar K^*\ell^+\ell^-$ 
provides a relation between the coefficient of the magnetic 
penguin operator $C_7$ and $C_9$ which is nearly free of 
hadronic uncertainties \cite{Ali:1999mm}:
\begin{equation} 
\label{afbrel}
 C_9 + \mbox{Re}(Y(q_0^2)) 
 =  - \frac{2 M_B m_b}{q_0^2} \, C_7^{\rm eff},
\end{equation}
where $Y$ is a calculable function. The next-to-leading order 
corrections to the exclusive decay have recently been 
calculated \cite{Beneke:2001at} (using the new result for the 
two-loop 
virtual correction to the inclusive decay \cite{Asatrian:2001de}) 
and have been shown to correct Eq.~(\ref{afbrel}) significantly, 
as seen from the shift of the asymmetry zero in 
Figure~\ref{fig:dafbnorm}. However, the conclusion 
that $C_9$ can be determined from the asymmetry zero with 
little uncertainty remains valid after this correction is taken into 
account.

\section{Conclusion}

The three sets of observables discussed here could easily have come 
out different from their Standard Model expectation, but apparently 
did not. The challenge is therefore to understand what distinguishes 
the Kobayashi-Maskawa mechansim of CP violation and the GIM mechanism 
of suppressing flavour-changing neutral currents from the many 
sources of CP violation and flavour-changing neutral currents that 
arise in generic extensions of the Standard Model.

\section{Acknowledgements}
I wish to thank the organisers for their generous support and for
organising a splendid conference.

\end{document}